\documentclass[usenatbib]{mn2e}

\usepackage{amsmath,amssymb, graphics}
\usepackage{pgf,tikz}
\usepackage{bm}
\usepackage{graphicx}
\usepackage{epstopdf}

\newcommand{\Inert}{{\bf I}}
\newcommand{\Mag}{{\bf M}}
\newcommand{\T}{{\bf T}}
\newcommand{\one}{{\bm \delta}}

\newcommand{\Ieone}{I_{\text{eff},1}}
\newcommand{\Ietwo}{I_{\text{eff},2}}
\newcommand{\Iethree}{I_{\text{eff},3}}
\newcommand{\Dip}{{\bf M}_P}
\newcommand{\Quad}{{\bf M}_Q}
\newcommand{\vomega}{{\bm \omega}}
\newcommand{\vGamma}{{\bm \Gamma}}
\newcommand{\vp}{{\bf p}}
\newcommand{\ve}{{\bf e}}
\newcommand{\vL}{{\bf L}}
\newcommand{\der}{\text{d}}
\newcommand{\cn}{\text{cn}}
\newcommand{\sn}{\text{sn}}
\newcommand{\dn}{\text{dn}}

\newcommand{\eoneeff}{\hat {\bf e}_{\text{eff},1}}
\newcommand{\etwoeff}{\hat {\bf e}_{\text{eff},2}}
\newcommand{\ethreeeff}{\hat {\bf e}_{\text{eff},3}}
\newcommand{\eps}{\epsilon}

\newcommand{\vdel}{{\bm \nabla}}
\newcommand{\E}{{\bf E}}
\newcommand{\B}{{\bf B}}
\newcommand{\vv}{{\bf v}}
\newcommand{\br}{{\bf r}}
\newcommand{\vY}{{\bf Y}}
\newcommand{\bQ}{{\bf Q}}
\newcommand{\p}{\hat{\bf p}}
\newcommand{\qone}{\hat{\bf q}_1}
\newcommand{\qtwo}{\hat{\bf q}_2}
\newcommand{\qthree}{\hat{\bf q}_3}
\newcommand{\bcdot}{{\bm \cdot}}
\newcommand{\btimes}{{\bm \times}}
\newcommand{\vPhi}{{\bm \Phi}}
\newcommand{\vPsi}{{\bm \Psi}}
\newcommand{\xo}{x_0}
\newcommand{\im}{\textit{i}}

\newcommand{\executeiffilenewer}[3]{%
 \ifnum\pdfstrcmp{\pdffilemoddate{#1}}%
 {\pdffilemoddate{#2}}>0%
 {\immediate\write18{#3}}\fi%
}

\newcommand{\be}{\begin{equation}}
\newcommand{\ee}{\end{equation}}

\title[Electromagnetic Torques, Precession and Evolution of Magnetic Inclination of Pulsars]{Electromagnetic Torques, Precession and Evolution of Magnetic Inclination of Pulsars}
\author[J. J. Zanazzi and Dong Lai]{J. J. Zanazzi$^{1}$\thanks{Email: jjz54@cornell.edu}, and Dong Lai$^{1}$ \\
$^{1}$Center for Space Research, Department of Astronomy,
Cornell University, Ithaca, New York 14853}

\begin{document}

\maketitle
\begin{abstract}
We present analytic calculations of the electromagnetic torques acting on a magnetic neutron star rotating in vacuum, including near-zone torques associated with the inertia of dipole and quadrupole magnetic fields.  We incorporate these torques into the rotational dynamics of a rigid-body neutron star, and show that the effects of the inertial torque can be understood as a modification of the moment of inertia tensor of the star.  We apply our rotational dynamics equation to the Crab pulsar, including intrinsic distortions of the star and various electromagnetic torques, to investigate the possibility that the counter-alignment of the magnetic inclination angle, as suggested by recent observations, could be explained by pulsar precession.  We find that if the effective principal axis of the pulsar is nearly aligned with either the magnetic dipole axis or the rotation axis, then precession may account for the observed counter-alignment over decade timescales.  Over the spindown timescale of the pulsar, the magnetic inclination angle always decreases.
\end{abstract}


\section{Introduction}

The structure and evolution of magnetic fields is one of the key
ingredients to understanding various observational manifestations of
radio pulsars and other types of neutron stars (NSs) (e.g., \citealt{HardingLai(2006), Kaspi(2010), Reisenegger(2013)}).
For radio pulsars, the magnetic inclination angle $\alpha$, defined as the angle
between the pulsar's magnetic dipole axis and rotation axis, strongly
affects the pulse and polarization profiles in radio and high energy
emissions (e.g., \citealt{Rookyard(2015)}).
By analyzing polarization data for a large number of pulsars, \cite{TaurisManchester(1998)} found that, statistically, pulsars with large
characteristic ages tend to have small magnetic inclination angles,
suggesting that the magnetic axis align with the spin axis on a
timescale of order $10^7$~years (see \citealt{WeltevredeJohnston(2008)}
and \citealt{Young(2010)}, who found somewhat different alignment timescales).
On the other hand, general pulsar population studies have revealed no
evidence for significant torque decay (due to magnetic field decay or
alignment) over the pulsar lifetime ($\sim 10^8$~years) (e.g.,
\citealt{Faucher(2006),Gullon(2014)}).

Recently, \cite{Lyne(2013)} found that the radio pulse profile of the
Crab pulsar has shown a steady increase in the separation of the main
pulse and interpulse components at $0.6^\circ$ per century over 22
years (see also \citealt{Lyne(2015)}).
The increase in pulse seperation was interpreted as an increase in the magnetic inclination angle $\alpha$.
(see \citealt{Watters(2009)}).
This interpretation is also consistent with departure of the braking
index $n = (\omega \ddot \omega)/\dot \omega^2$ from 3 for the Crab
pulsar (where $\omega$ is the angular rotation frequency). Using 
the braking torque due to a rotating magnetic dipole in vacuum,
$\dot\omega\propto -\omega^3\sin^2\alpha$, and assuming a constant magnetic dipole
moment, the braking index is given by 
\be\label{eq:braking}
n = 3 + 2 \frac{\omega}{\dot \omega} \frac{\dot \alpha}{\tan \alpha}.
\ee
With the observed $\dot \alpha = 0.6^\circ/\text{century}$ and $\omega/\dot\omega= -24.9 \, \text{century}$, and the estimate $\alpha \approx 45^\circ$ \citep{Harding(2008)}, we find $n \approx 2.48$,
in agreement with the observed value of $n \simeq 2.50$ \citep{Lyne(2013)}.


The increase in the magnetic inclination angle cannot be explained by
the simplest dynamical model of neutron stars (NSs).  If one models a
NS as a spherical body endowed with a frozen-in dipole magnetic field
in vacuum, one expects only a decrease in the magnetic
inclination angle \citep{DavisGoldstein}.  Including the electro-dynamical effects of
the magnetosphere leads to pulsar spindown even when the NS has an
aligned dipole field ($\alpha = 0$) \citep{Spitkovsky(2006), KalapotharakosContopoulos(2009), Kalapotharakos(2012), Tchekhovskoy(2013)},
but still predicts magnetic alignment with the rotation axis
\citep{Philippov(2014)}.  
While these results may be consistent with pulsar population statistics
\citep{TaurisManchester(1998)}, the short-term ($\sim$10 years) increase
of $\alpha$ observed in the Crab pulsar is unaccounted for.


A possible physical mechanism for magnetic counter-alignment (increase
of $\alpha$) is pulsar precession, a topic of interest for nearly half
a century.  Early models of free
precession modeled the NS
as a rigid body undergoing a torque due to angular
momentum loss from dipole radiation \citep{DavisGoldstein, Goldreich}. Then came the inclusion of the
pinned superfluid in the NS crust, which was shown to
severely alter the rotational dynamics of NSs \citep{Shaham(1977), Alpar(1984)1, AlparOegelman(1987), CasiniMontemayor(1998), Sedrakian(1999), LinkCutler}.  The effects of super-fluidity ``destroyed"
precession, speeding it up to a rate undetectable by observations.
But tentative observational evidence suggested that some pulsars precessed with periods
comparable to those predicted by free precession
\citep{SutoIso(1985), Truemper(1986), Lyne(1988), Weisberg(2010), Makishima(2014)}.  This led many to still model the precession
of NSs as free, rather than forced
\citep{LinkEpstein(1997), MelatosCurrentStarved, MelatosBumpySpin, MelatosRadPrecess, JonesAndersson(2001), LinkEpstein(2001), Wasserman(2003)}, and to infer interior physics which would lead to weak coupling between the
crust and the core.  Overall, despite the uncertainties, free precession remains a possible
model for understanding the rotational behavior of NSs
\citep{Jones(2012)}.

In this paper, we treat the NS as a non-spherical rigid body acted upon by electromagnetic (EM) torques.  Section \ref{sec:torque} presents our calculation of the EM torques, including both ``inertial torques" associated with the inertia of the near-zone EM field, and dipole radiative torque.  In section
\ref{sec:dynamics}, we solve the equations of motion for the NS rotation analytically, reproducing results which before now were only
studied numerically \citep{MelatosBumpySpin, MelatosRadPrecess}.  We show that the main effects of the inertial torque may be understood by modifying the moment of inertia tensor of the NS.  In section \ref{sec:applications}, we
discuss applications to pulsars, and in particular to the observed magnetic inclination evolution of the Crab
pulsar.  In section \ref{sec:Conclusion}, we summarize our findings, and discuss various uncertainties and possible future works.

\section{EM Torques on Rotating NS in Vacuum}
\label{sec:torque}

In this section, we calculate the EM torques on a
rotating, magnetized sphere in vacuum.  We consider both dipole and
quadrupole magnetic field topologies.  It is well known that a rotating
magnetic NS must be surrounded by a magnetosphere with current and charge distributions
\citep{GoldreichJulian(1969)}.  This magnetosphere modifies the
magnetic breaking torque significantly
\citep{Spitkovsky(2006), KalapotharakosContopoulos(2009), Philippov(2014)}.  
However, the torque associated with the near-zone magnetic field inertia has not
been calculated for magnetosphere models.  We will show this inertial torque can significantly 
affect the precession dynamics of the NS. 

\subsection{Dipole Field}

A spherical body with endowed with a dipole field rotating in
vacuum has two torques acting on it.  The first arises from the fact
that a misaligned spinning dipole emits EM radiation, carrying away
angular momentum.  We denote this torque as 
$\vGamma_\text{rad}$.  
The second torque arises from the inertia of the dipole
magnetic field \citep{DavisGoldstein}, which we will denote by $\vGamma_P$.  Our 
calculation yields the expressions (see Appendix A)
\be\label{eq:RadTorque}
\vGamma_\text{rad} = \frac{2 \omega^2}{3 c^3} (\vomega \btimes \vp) \btimes \vp,
\ee
\be\label{eq:DipInertTorque}
\vGamma_P = \frac{3}{5Rc^2} (\vp \bcdot \vomega)(\vp \btimes \vomega),
\ee
where $\vomega$ is the rotation rate vector, $\vp$ is the dipole
moment, $p = |\vp| = B_P R^3/2$, with $B_P$ the magnitude of the
fields at the magnetic poles, and $R$ is the radius of the NS.  

Note that the numerical coefficient $3/5$ in front of the expression for $\vGamma_P$
agrees with \cite{MelatosCurrentStarved}, but
disagrees with \cite{DavisGoldstein}, \cite{Goldreich}, \cite{GoodNg}, and \cite{Beskin(2013)}, all of whom quoted slightly different values.   Our equation \eqref{eq:DipInertTorque} is obtained by assuming a uniform interior field $\B_P$ which rotates rigidly around the spin axis, and an electric field given by $\E~=~-~(\vv/c)~\btimes~\B_P$.   Although this interior EM field was assumed, equation \eqref{eq:DipInertTorque} only depends on the exterior EM field on the surface of the NS [see Eq. \eqref{eq:TorqueComps} in Appendix].

The difference between our value and that given by 
\cite{GoodNg} and \cite{Beskin(2013)} may be attributed to the method used to calculate the torque.  These authors obtained the torque directly through the volume integral
\be
\vGamma = \int \br \btimes \left(\rho_e \E + \frac{1}{c} {\bf j}_e \btimes \B \right) \der V ,
\ee
and adopted specific assumptions on the charge density $\rho_e$ and the current density ${\bf j}_e$ inside the NS.  \cite{Beskin(2013)} also included the effects of the EM field's inertia.
The difference from the value of \cite{DavisGoldstein} cannot be
attributed to such a difference, as the authors appeared to have used the same method to calculate the torque. 
\cite{BeskinZheltoukhov(2014)} also obtained the coefficient $3/5$ using
the same approach as ours, although they questioned its validity
of the method, suggesting that the final answer depends on the
internal current distribution inside the rotating NS.  Indeed, the angular momentum carried by the EM field is
\be
{\bf L}_\text{EM} = \frac{1}{4 \pi c} \int \br \btimes (\E \btimes \B) \, \der V.
\ee
The rate of change of ${\bf L}_\text{EM}$ due to stellar rotation is of order
\be
\Omega |{\bf L}_\text{EM}| \sim \frac{B_P^2 R^5 \Omega^2}{c^2} \sim \frac{p^2 \Omega^2}{Rc^2},
\ee
the same order as equation \eqref{eq:DipInertTorque}.  Thus, including ${\bf L}_\text{EM}$ into the angular momentum equation amounts to a modification of the inertial torque expression \eqref{eq:DipInertTorque} by a factor of order unity [see Eq. \eqref{eq:Euler} of section \ref{sec:PrecSol}].  In the remainder of this paper, we adopt equation \eqref{eq:DipInertTorque} as the inertial torque and do not include ${\bf L}_\text{EM}$ into the dynamical equation for the NS.  Since the precise value of $B_P$ is uncertain by at least a factor of two, and more importantly, since the NS suffers a much larger deformation than that associated with the field inertia (see section \ref{sec:deformations}), a correction to equation \eqref{eq:DipInertTorque} by a factor of order unity does not affect the main results of our paper (see sections \ref{sec:dynamics}-\ref{sec:applications}).


For convenience, we define the dimensionless parameter $\eps_P$ as
\be\label{eq:epsP}
\eps_P \equiv \frac{3}{20}\frac{B_P^2 R^5}{I c^2},
\ee
where $I$ is the moment of inertia of NS.  With this
definition, $\Gamma_P$ becomes
\be\label{eq:InertTorque}
\vGamma_P = I \, \eps_P (\hat \vp \bcdot \vomega)(\hat \vp \btimes \vomega),
\ee
where $\hat \vp = \vp/p$ is the unit vector along the dipole axis.  Thus, a change of the coefficient in equation \eqref{eq:DipInertTorque} amounts to a modification of the value of $\eps_P$ by a factor of order unity.

\subsection{Quadrupole Field}

\begin{figure}
    \centering
    \includegraphics[scale=0.5]{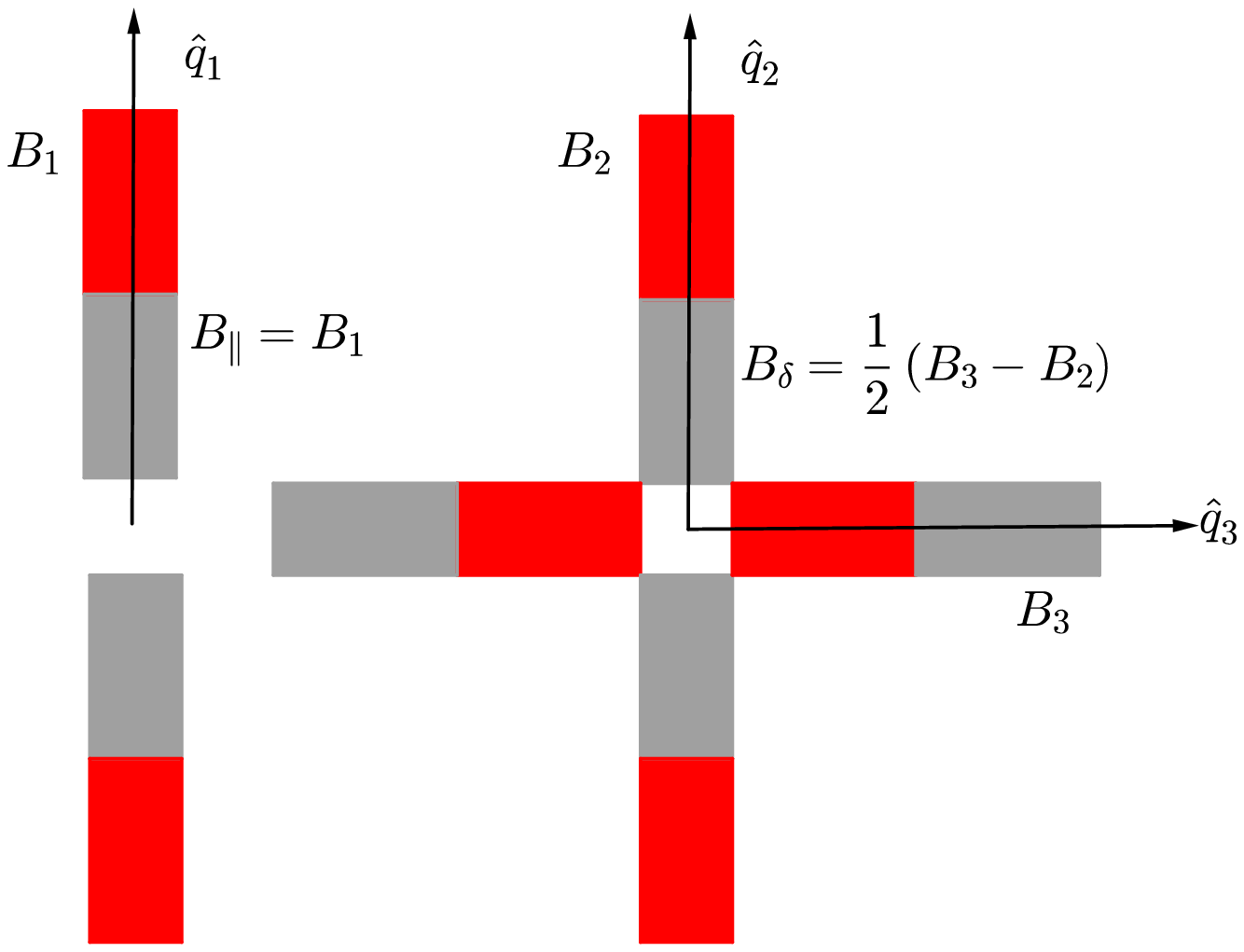}
    \caption{Two components of the quadrupole
      field.  A general quadrupole field is specified by three
      principal axes $( \qone, \qtwo, \qthree)$, in which the
      quadrupole tensor is diagonal, and two elements ($Q_\parallel$
      and $Q_\delta$) with the corresponding surface magnetic fields
      $B_\parallel$ and $B_\delta$.}
    \label{fig:Quad_field}
\end{figure}

We have also calculated the inertial torque for NSs with an arbitrary
magnetic quadrupole moment $\bQ$, where $\bQ$ is a symmetric
trace-free tensor with eigenvectors $\qone$, $\qtwo$, and $\qthree$,
and eigenvalues $Q_1$, $Q_2$, and $Q_3$ (see Appendix). Before we
state the result, we will explain our field decomposition.  Because $\bQ$ is
trace-free ($\sum_i Q_i = 0$), this quadrupole moment may be expressed
as the sum of two linearly independent tensors:
$\bQ_\parallel$ and $\bQ_\delta$.  These two independent trace-free quadrupole tensors 
have components in the $\{\hat {\bf q}_i \}$ basis given by
\begin{align}
\bQ_\parallel &= 
\begin{pmatrix}
-Q_\parallel/2 & 0 & 0 \\
0 & -Q_\parallel/2 & 0 \\
0 & 0 & Q_\parallel
\end{pmatrix},
\\
\bQ_\delta &= 
\begin{pmatrix}
Q_\delta & 0 & 0 \\
0 & -Q_\delta & 0 \\
0 & 0 & 0
\end{pmatrix},
\end{align}
where $Q_\parallel = Q_3$ and $Q_\delta = (Q_1 - Q_2)/2$.  On
the surface of the star $\br = R \hat \br$, the radial components
of the magnetic fields are
\begin{align}
\hat \br \bcdot \B_\parallel(R\hat \br) = &B_\parallel P_2(\hat \br \bcdot \qthree),  \\
\hat \br \bcdot \B_\delta(R \hat \br) = &\frac{2 B_\delta}{3} \left[ P_2(\hat \br \bcdot \qtwo) - P_2(\hat \br \bcdot \qone) \right],
\end{align}
where $B_\parallel = -2Q_\parallel/R^4$ and $B_\delta = 3
Q_\delta/R^4$, and $P_2$ denotes the Legendre polynomial of order 2.  From this,
we see that the magnetic field of a general quadrupole is completely
specified by the basis vectors $\{ \hat {\bf q}_i\}$, and the
field strengths $B_\parallel$ and $B_\delta$.  Figure
\ref{fig:Quad_field} illustrates the geometry of these field components.

The above composition of the surface quadrupole magnetic field completely specifies the external EM fields, which give rise to an inertial torque (see Appendix for calculation)
\begin{align}\label{eq:QuadInertTorque}
\vGamma_Q &= \frac{B_\parallel^2 R^5}{175 c^2} (\vomega \bcdot \qone)(\vomega \btimes \qone) \nonumber \\
+ &\frac{4 B_\parallel B_\delta R^5}{525 c^2} \left[ (\vomega \bcdot \qthree)(\vomega \btimes \qthree) - (\vomega \bcdot \qone)(\vomega \btimes \qone) \right] \nonumber \\
- &\frac{4 B_\delta^2 R^5}{1575 c^2} \left[ (\vomega \bcdot \qthree)(\vomega \btimes \qthree) + (\vomega \bcdot \qtwo)(\vomega \btimes \qtwo) \right].
\end{align}
The form of this torque differs from \cite{GoodNg}. The torque calculated in \cite{GoodNg} is expressed in Cartesian coordinates, with $\hat \vomega = \hat z$, while the expression above is basis independent.
For convenience, we define two new dimensionless parameters:
\begin{align}
\eps_\parallel &\equiv \frac{1}{175} \frac{B_\parallel^2 R^5}{Ic^2}, \\
\eps_\delta &\equiv \frac{4}{1575} \frac{B_\delta^2 R^5}{I c^2}.
\end{align}
With these definitions, the inertial magnetic quadupole torque takes the form
\begin{align}\label{eq:QtorquePaper} 
\vGamma_Q &= I \, \eps_\parallel(\vomega \bcdot \qone)(\vomega \btimes \qone) \nonumber \\
+ &2 I \sqrt{\eps_\parallel \eps_\delta} \left[ (\vomega \bcdot \qthree)(\vomega \btimes \qthree) - (\vomega \bcdot \qone)(\vomega \btimes \qone) \right] \nonumber \\
- &I \, \eps_\delta  \left[ (\vomega \bcdot \qthree)(\vomega \btimes \qthree) + (\vomega \bcdot \qtwo)(\vomega \btimes \qtwo) \right]. 
\end{align}

\subsection{Magnetic Inertia Tensor}

To re-express the torques given by equations \eqref{eq:InertTorque}
and \eqref{eq:QtorquePaper}, we define two tensors $\Dip$ and $\Quad$,
associated with the inertia of the dipole and quadrupole magnetic
fields:
\begin{align}
\Dip &\equiv -I \, \eps_P (\p \otimes \p), \label{eq:Mp} \\
\Quad &\equiv - I \, \eps_\parallel \left( \qone \otimes \qone \right) \nonumber \\
- &2 I \sqrt{ \eps_\parallel \eps_\delta } \left( \qthree \otimes\qthree - \qone \otimes \qone \right) \nonumber  \\
+ &I \, \eps_\delta \left( \qthree \otimes \qthree + \qtwo \otimes \qtwo \right) \label{eq:Mq}.
\end{align}
The Magnetic Inertia Tensor $\Mag$ is defined to be
\be\label{eq:M}
\Mag \equiv \Dip + \Quad,
\ee
so that the inertial torque takes the form
\be\label{eq:MagInert}
\vGamma_\text{inert} = \vGamma_P + \vGamma_Q = -\vomega \btimes (\Mag \bcdot \vomega).
\ee
This new form of the inertial torque $\vGamma_\text{inert}$ given by
equation~(\ref{eq:MagInert}) will be used in the next section.

\section{Non-Dissipative Precession}
\label{sec:dynamics}

In this section, we solve the equation
of motion for the NS rotation analytically, incorporating a non-spherical rigid body moment of inertia tensor and the
inertial torques from co-rotating dipole and quadrupole magnetic fields.
We neglect the radiative torque, and derive an analytic expression for precession period.  \cite{MelatosBumpySpin, MelatosRadPrecess} has previously presented numerical solutions for such non-dissipative precision, incorporating only the dipole torque.

\subsection{Explicit Solution for Non-Dissipative Precession}
\label{sec:PrecSol}

The Euler equation for the NS rotation takes the form
\be\label{eq:Euler}
\Inert \bcdot \frac{\der \vomega}{\der t} + \vomega \btimes \left( \Inert \bcdot \vomega \right) = \vGamma_\text{inert},
\ee
where $\Inert$ is the ``intrinsic" moment of inertia tensor for the NS.  The derivative $\der/ \der t$ is taken in the frame with the body.  Substituting expression \eqref{eq:MagInert} into the dynamical equation \eqref{eq:Euler} and re-arranging gives
\be\label{eq:ExactDiffyq}
\Inert \bcdot \frac{\der \vomega}{\der t} + \vomega \btimes \left[ \left( \Inert + \Mag\right) \bcdot \vomega \right] = 0.
\ee
To solve this system analytically, we define an effective inertia tensor $\Inert_\text{eff}$ as
\be\label{eq:InertEff}
\Inert_\text{eff} \equiv \Inert + \Mag.
\ee
As long as the magnitude $|\Mag|/I$ is much less than unity, or $\eps_P,\eps_\parallel,\eps_\delta \ll 1$, we can replace the first term in \eqref{eq:ExactDiffyq} by $\Inert_\text{eff} \bcdot \der \vomega/\der t$.  This approximation is valid in the full regime of interest [see Eq. \eqref{eq:DipMag}, \eqref{eq:QparMag}, and \eqref{eq:QdeltaMag} below].  Equation \eqref{eq:ExactDiffyq} then becomes
\be\label{eq:ApproxDiffyq}
\frac{\der \vL}{\der t} + \vomega \btimes \vL= 0,
\ee
where $\vL \equiv \Inert_\text{eff} \bcdot \vomega$ is the effective angular momentum of the body, including the inertia term from the near-zone magnetic field.

Equation \eqref{eq:ApproxDiffyq} is the equation of motion for a freely precessing rigid body, and has a well known analytic solution which we will summarize.  Because $\Inert_\text{eff}$ is a real $3\times 3$ symmetric tensor, it may be diagonalized.  Let $\ethreeeff$, $\etwoeff$, and $\eoneeff$ denote the three eigenvectors (the principal axes) of $\Inert_\text{eff}$ with the respective eigenvalues $\Iethree > \Ietwo > \Ieone$.  In this frame, equation \eqref{eq:ApproxDiffyq} has the conserved quantities
\begin{align}\label{eq:ConservedQuantities}
\frac{L_1^2}{\Ieone} + \frac{L_2^2}{\Ietwo} + \frac{L_3^3}{\Iethree} &= 2E, \\
L_1^2 + L_2^2 + L_3^2  &= L^2,
\end{align}
where $E$ is the rotational energy of the body and $L$ is the angular momentum.  The evolution of the components of $\hat \vL \equiv \vL/L$ can be obtained by solving equation \eqref{eq:ApproxDiffyq} (see \citealt{LLMechanics} Chapter 6; \citealt{AkgunLinkWasserman}):
\begin{equation}\label{eq:sol1}
\begin{array}{cl}
\hat L_1 &= -\Lambda \; \cn(\phi,k^2), \\
\hat L_2 &= -\Lambda \sqrt{1+e^2} \; \sn(\phi,k^2), \\
\hat L_3 &= \sqrt{1-\Lambda^2} \; \dn(\phi,k^2),
\end{array}
\end{equation}
when $L^2>2E \Ietwo$ and
\begin{equation}\label{eq:sol2}
\begin{array}{cl}
\hat L_1 &= -\Lambda \, \dn(k\phi,k^{-2}), \\
\hat L_2 &= -\sqrt{(1-\Lambda^2)(1+e^{-2})} \, \sn(k\phi,k^{-2}), \\
\hat L_3 &= \sqrt{1-\Lambda^2} \, \cn(k \phi,k^{-2}),
\end{array}
\end{equation}
when $L^2<2E \Ietwo$.  Here $\cn$, $\sn$, and $\dn$ are the Jacobian Elliptic functions, and
\begin{align}
\Lambda &= \sqrt{ \frac{\Ieone(2E\Iethree- L^2)}{L^2(\Iethree-\Ieone)} }, \\
\phi &= t \, \omega_p, \\
\omega_p &= \frac{\epsilon_\text{eff} L \sqrt{1-\Lambda^2}}{\Iethree \sqrt{1+e^2}}, \\
k^2 &= \frac{e^2 \Lambda^2}{1-\Lambda^2}, \\
\epsilon_\text{eff} &= \frac{\Iethree - \Ieone}{\Ieone}, \\
e^2 &= \frac{\Iethree(\Ietwo-\Ieone)}{\Ieone(\Iethree-\Ietwo)}.
\end{align}
Equations \eqref{eq:sol1} imply precession around $\ethreeeff$, and equations \eqref{eq:sol2} imply precession around $\eoneeff$.  This shows the main effect of the inertial torque is to modify the equations of motion from that of a freely-precessing body with moment of inertia $\Inert$ to a freely-precessing body with a modified moment of inertia $\Inert_\text{eff}$.

An effectively biaxial body corresponds to the special cases of $e = 0$ or $e = \infty$.  When $e = 0$, equations \eqref{eq:sol1} simplify to
\begin{equation}\label{eq:sol1bi}
\begin{array}{cl}
\hat L_1 &= -\sin\theta \, \cos\phi,\\
\hat L_2 &= - \sin\theta \, \sin\phi, \\
\hat L_3 &= \cos\theta,
\end{array}
\end{equation}
where
\begin{equation}
\begin{array}{rcl}
\phi &= &\cos\theta \, \eps_\text{eff} \, \omega \, t \\
\end{array}
\end{equation}
with $\Lambda = \sin\theta$ and $\cos\theta = \hat \bomega \bcdot \ethreeeff$.  When $e = \infty$, equations \eqref{eq:sol2} simplify to
\begin{equation}\label{eq:sol2bi}
\begin{array}{cl}
\hat L_1 &= -\cos\theta, \\
\hat L_2 &= -\sin\theta \, \sin(k\phi), \\
\hat L_3 &= \sin\theta \, \cos(k \phi).
\end{array}
\end{equation}
where
\begin{equation}
\begin{array}{rcl}
k\phi &= &\cos\theta \, \eps_\text{eff} \, \omega \, t \\
\end{array}
\end{equation}
and $\Lambda = \cos\theta = \hat \bomega \bcdot \eoneeff$.

\subsection{Numerical examples}

\begin{figure}
    \centering
    \includegraphics[scale=0.4]{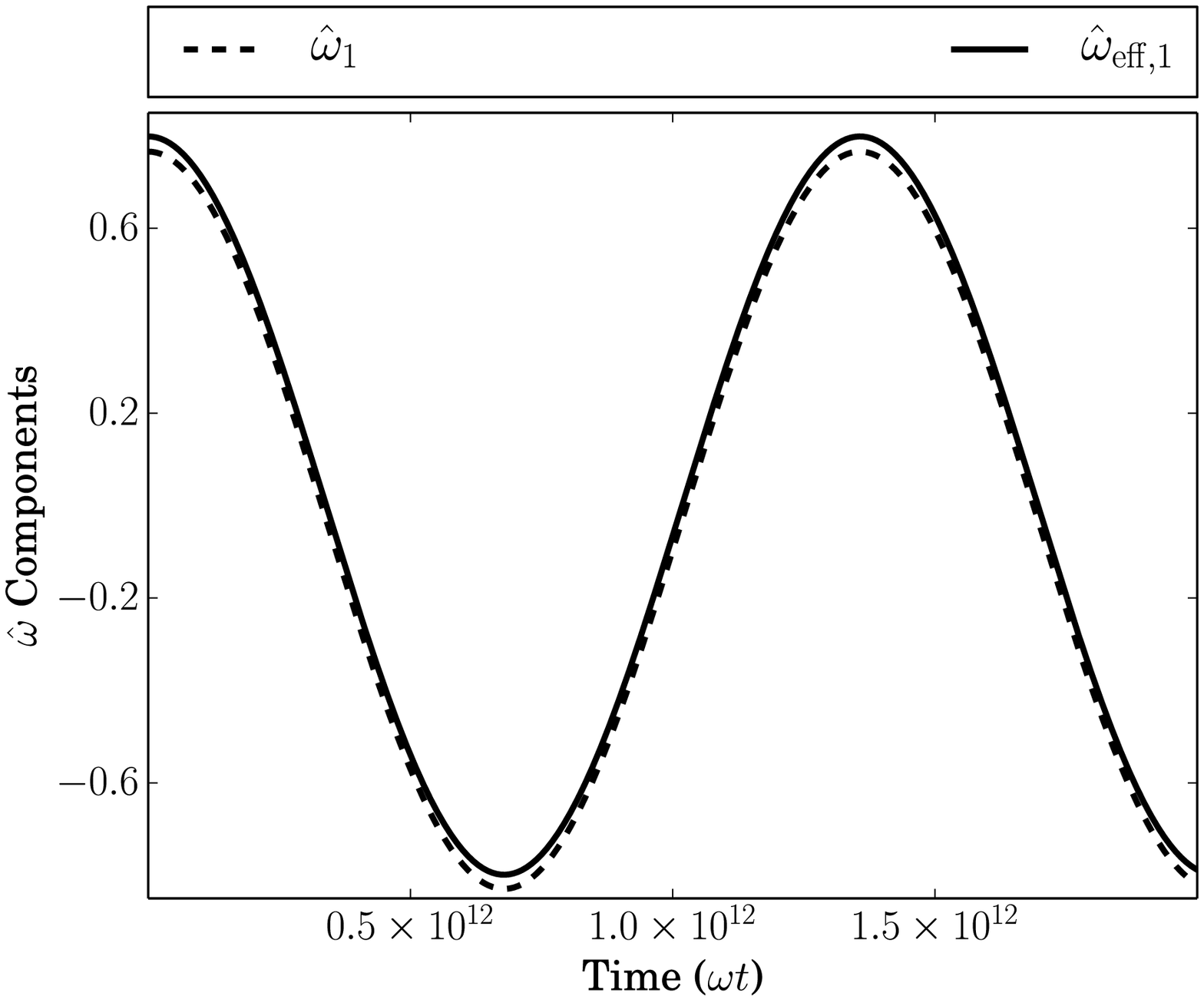}
    \includegraphics[scale=0.4]{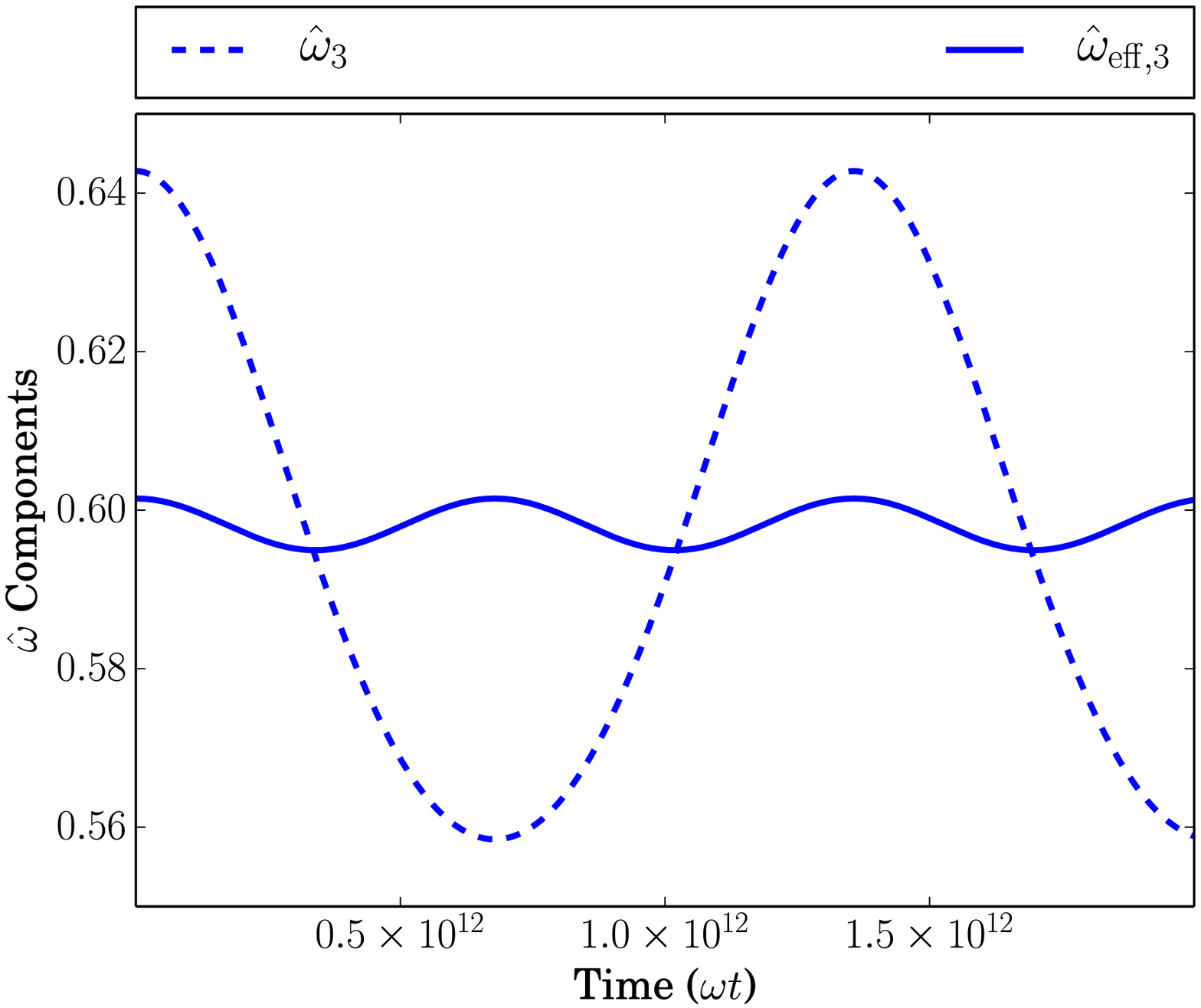}
    \caption{Components of the angular velocity unit vector $\hat \vomega = \vomega/\omega$ obtained by integrating equation \eqref{eq:ExactDiffyq}.  The components are $\hat \omega_i = \hat \ve_i \bcdot \hat \vomega$ (dashed line) and $\hat \omega_{\mathrm{eff},i} = \hat \ve_{\mathrm{eff},i} \bcdot \hat \vomega$ (solid line). We assume a biaxial star with ellipticity $\epsilon = 10^{-11}$, and $\p$ oriented in the $(13)$-plane with angle $\chi=10^\circ$, and $\epsilon_P=2.4\times 10^{-12}$.  We assume that $\epsilon_Q = 0$.  In the $\{ \hat \ve_i \}$ basis, $\hat \ve_1 = (1,0,0)$, $\eoneeff = (0.9986,0,0.05277)$, $\hat \ve_3 = (0,0,1)$ and $\ethreeeff = (-0.05277,0,0.9986)$. Note $\hat \ve_2 = \etwoeff$.  The initial conditions are $\hat \ve_1 \bcdot \hat \vomega = \sin(50^\circ)$, $\hat \ve_2 \bcdot \hat \vomega = 0$, and $\hat \ve_3 \bcdot \hat \vomega(0) = \cos(50^\circ)$.  The fact that $\ethreeeff \bcdot \hat \vomega \simeq \text{const.}$ indicates that $\hat \vomega$ is precessing around $\ethreeeff$, and the small variation is due to the fact that $\Inert_\text{eff}$ is slightly triaxial.}
\label{fig:Prec_Dipole}
\end{figure}

\begin{figure}
    \centering
    \includegraphics[scale=0.4]{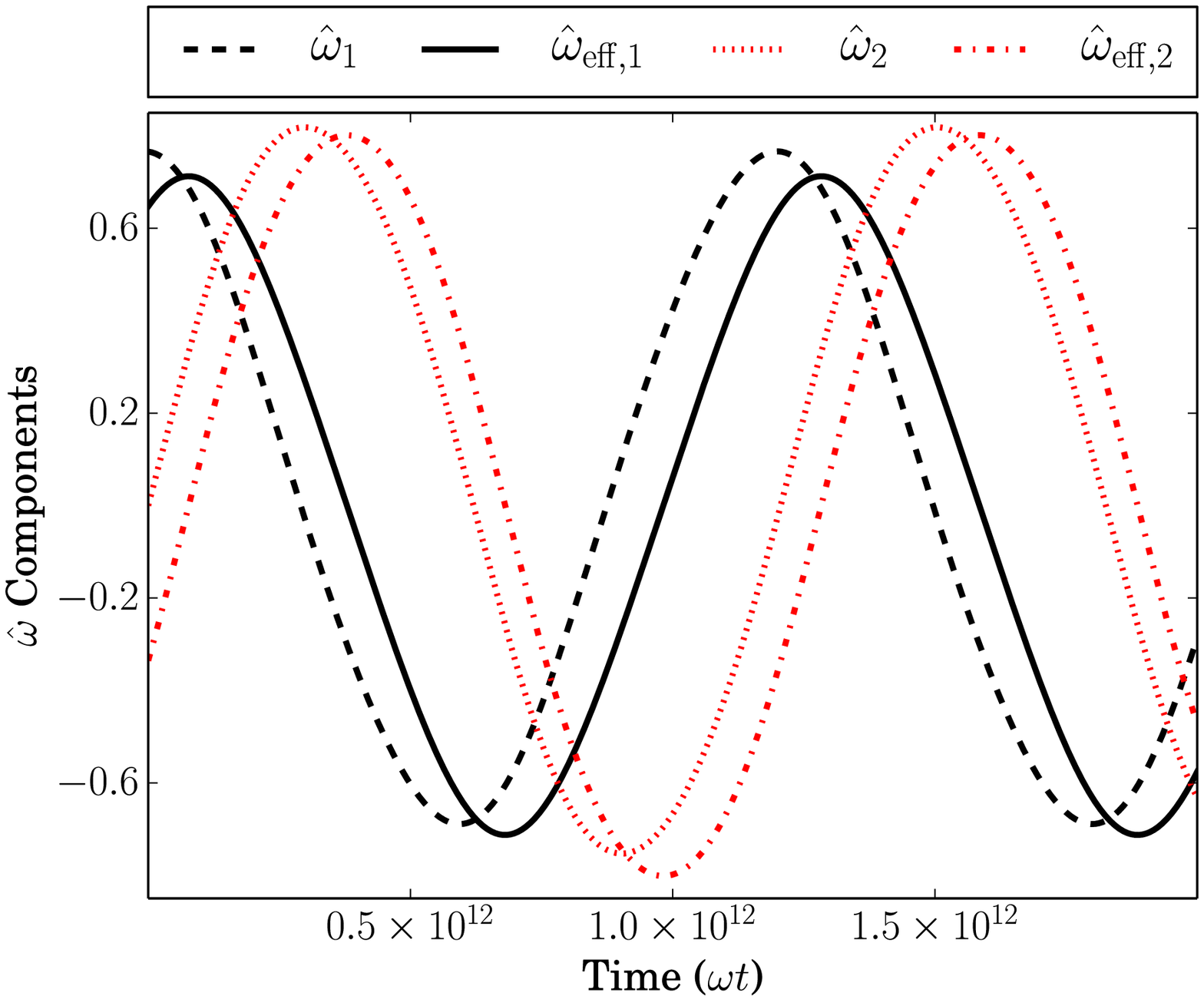}
    \includegraphics[scale=0.4]{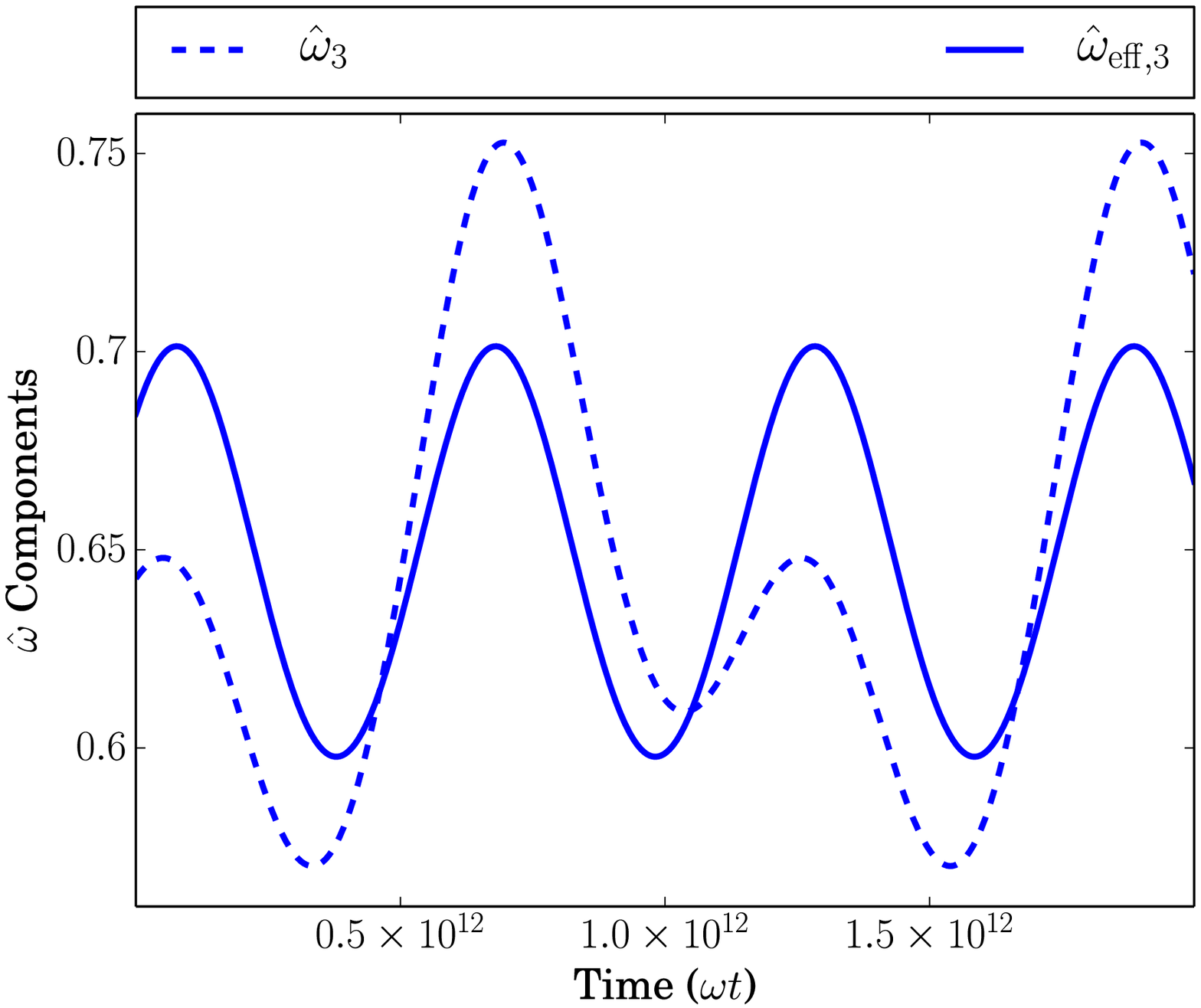}
    \caption{Same as Figure \ref{fig:Prec_Dipole}, except we have added a quadrupole inertial torque with $\epsilon_\parallel = \epsilon_P$, $\epsilon_\delta = 0$, and $\qone = [\sin(60^\circ)\cos(205^\circ),\sin(60^\circ)\sin(205^\circ),\cos(60^\circ)]$ in the basis $\{\hat \ve_i\}$.  Because $\eps_\delta = 0$, $\qtwo$ and $\qthree$ are irrelevant.  Notice that $\ethreeeff \bcdot \hat \vomega$ is no longer constant, because the tensor $\Inert_\text{eff}$ is much more tri-axial.}
\label{fig:Prec_Quad}
\end{figure}

In this subsection, we illuminate the solutions of equation \eqref{eq:ApproxDiffyq} with some illustrative examples.  We define a body axis $\{ \hat \ve_i \}$, corresponding to the eigenvectors of the axis-symmetric tensor $\Inert$, with an intrinsic ellipticity $\eps \equiv (I_3 - I_1)/I_3$ and a dimensionless angular velocity parameter $\hat \vomega \equiv \vomega/\omega$.
Figures \ref{fig:Prec_Dipole} and \ref{fig:Prec_Quad} show the time evolution of $\hat \vomega$.  In both figures, an intrinsically biaxial star is assumed, with $\epsilon = 10^{-11}$ (see sec. \ref{sec:deformations}) and the symmetry axis along $\hat \ve_3$.

Figure \ref{fig:Prec_Dipole} shows how the dynamics are modified by a dipole magnetic field with $\epsilon_P = 2.4 \times 10^{-12}$ and direction $\hat \vp$ oriented in the $(13)$ plane with angle $\chi=10^\circ$.  The fact that $\ethreeeff \bcdot \hat \vomega \simeq \text{const.}$ indicates that $\hat \vomega$ is precessing around $\ethreeeff$, with slight variations due to the fact that $\Inert_\text{eff}$ is slightly triaxial.  Figure \ref{fig:Prec_Quad} adds a quadrupole magnetic field with $\epsilon_\parallel = \epsilon_P$, $\epsilon_\delta = 0$, and $\qone = [\sin(60^\circ)\cos(205^\circ),\sin(60^\circ)\sin(205^\circ),\cos(60^\circ)]$.  Notice that $\ethreeeff \bcdot \hat \vomega$ is no longer constant, which follows from a more triaxial $\Inert_\text{eff}$ than the example displayed in Figure \eqref{fig:Prec_Dipole}.

The solutions to equation \eqref{eq:ApproxDiffyq} are periodic in time, with periods $T_{1}$ and $T_{3}$ for precession around $\eoneeff$ and $\ethreeeff$ respectively.  The explicit relation for $T_1$ and $T_2$ are given by
\begin{align}
T_1 &= \frac{4 \Ieone}{\epsilon_\text{eff} L} \sqrt{ \frac{1+e^{-2}}{1-\Lambda^2} } \frac{K(k^{-2})}{k}, \label{eq:T1} \\
T_3 &= \frac{4 \Iethree}{\epsilon_\text{eff} L} \sqrt{\frac{1+e^2}{1-\Lambda^2}} K(k^2), \label{eq:T3}
\end{align}
where $K$ is the complete elliptic integral of the first kind.  The reason why we do not have the simple relation $T = 2\pi/\omega_p$ is because the Jacobi Elliptic functions are not $2\pi$ periodic, but rather $4K(k^2)$ periodic.  For the special cases $e = 0$ or $e = \infty$, these relations simplify to
\begin{equation}
T_1 = T_3 = \frac{2 \pi}{\epsilon_\text{eff} \omega \cos\theta}.
\end{equation}
  One can show that the periods \eqref{eq:T1} and \eqref{eq:T3} give the periods displayed in Figures \ref{fig:Prec_Dipole} and \ref{fig:Prec_Quad}. 

\section{Application to Pulsars}
\label{sec:applications}

In the previous section, we showed that the effect of the inertial torque is to add an effective deformation in the moment of inertia of the NS.  In this section we apply this formalism to pulsars.  We begin with going over the relative magnitudes of NS deformations, then examine the behavior of the magnetic inclination angle over different timescales, and lastly discuss the possibility of explaining the magnetic counter-alignment of the Crab pulsar with precession.

\subsection{Neutron Star Deformations}
\label{sec:deformations}

There are several physical causes for the moment of inertia tensor $\Inert$ of a NS to depart from spherical symmetry.  The first is due to interior magnetic fields, which gives an intrinsic body ellipticity $\epsilon = (I_3-I_1)/I_1$ of order
\begin{multline}\label{eq:epsmag}
\epsilon_\text{mag} = \beta \frac{R^4 B_*^2}{GM^2} \\
= 2\times 10^{-12} \beta \left( \frac{B_*}{10^{12} \; \text{G}} \right)^2 \left( \frac{R}{10^6 \; \text{cm}}\right)^4 \left( \frac{M}{1.4 \; M_\odot} \right)^{-2},
\end{multline}
where $B_*$ is the internal magnetic field strength, and $\beta$ is a dimensionless constant which depends on the geometry of the internal field.  For a dipole or toroidal magnetic field topology, $\beta$ is of order unity \citep{Mastrano(2013)}.

The second source is rotation.  A uniform density fluid star rotating with angular velocity $\omega$ has an ellipticity of
\be
\eps_\text{fluid} = \frac{15}{16 \pi} \frac{\omega^2}{G\rho},
\ee
where $\rho$ is the density.  The NS is likely to have deformations of this order early in its lifetime.  Once the NS crust crystallizes, the body is able to support hydrostatic stresses.  If we idealize the crust as having a uniform shear modulus $\mu$, the part of the ellipticity which does not align with the rotation axis is \citep{MunkMacDonald}
\begin{multline}\label{eq:epselastic}
\eps_\text{elastic} = \frac{\tilde \mu}{1+\tilde \mu} \eps_\text{fluid} =   2\times 10^{-11} \left( \frac{\mu}{10^{30} \; \text{dynes/cm}^2} \right) \\
\times \left( \frac{P}{1 \; \text{sec.}} \right)^{-2} \left( \frac{R}{10^6 \; \text{cm}} \right)^7 \left( \frac{M}{1.4 \; M_\odot} \right)^{-3},
\end{multline}
where $\tilde \mu = 19 \mu/(2 \rho g R)$, $g$ is the surface gravity, and $\mu$ is the fiducial value for the shear modulus for the crust, evaluated at the density of order $\sim 10^{14} \, \text{g}/\text{cm}^3$.  This is a simple order of magnitude estimate of $\eps_\text{elastic}$.  More detailed calculations for realistic NS parameters may be found in \cite{Cutler(2003)}.

We compare these ``intrinsic" deformations in the moment of inertial $\Inert$ to the effective moments of inertia induced by co-rotating magnetic fields.  These are
\begin{multline}\label{eq:DipMag}
\epsilon_P = 1.5 \times 10^{-13} \left( \frac{B_P}{10^{12} \; \text{G}} \right)^2 \left( \frac{R}{10^6 \; \text{cm}} \right)^5 \left( \frac{M}{1.4 M_\odot}\right)^{-1},
\end{multline}
\begin{multline}\label{eq:QparMag}
\eps_\parallel = 5.7 \times 10^{-15}\left( \frac{B_\parallel}{10^{12} \; \text{G}} \right)^2 \left( \frac{R}{10^6 \; \text{cm}} \right)^5 \left( \frac{M}{1.4 M_\odot}\right)^{-1},
\end{multline}
\begin{multline}\label{eq:QdeltaMag}
\eps_\delta = 2.5 \times 10^{-15}\left( \frac{B_\delta}{10^{12} \; \text{G}} \right)^2 \left( \frac{R}{10^6 \; \text{cm}} \right)^5 \left( \frac{M}{1.4 M_\odot}\right)^{-1}.
\end{multline}
Thus, the biggest correction to the moment of inertia tensor $\Inert_\text{eff}$ comes from $\eps_P$, assuming that the quadrupole and dipole field strengths are similar.  In order for the corrections due to the quadrupole field to dominate over that of the dipole, one needs either $B_\parallel \gtrsim 5 B_P$ or $B_\delta \gtrsim 7 B_P$.

We note that $\epsilon_P$ is always at least an order of magnitude smaller than $\epsilon_\text{mag}$ (since $B_* \gtrsim B$), and is smaller than $\epsilon_\text{elastic}$ for realistic NS parameters. Of course, $\epsilon_P$ may be directly inferred from the measured $P$ and $\dot P$ of the pulsar:
\begin{equation}\label{eq:epsPPdot}
\epsilon_P \approx \frac{9}{40 \pi^2} \frac{c}{R} P \dot P.
\end{equation}
This gives an observational lower bound on the effective ellipticity $\eps_\text{eff}$ associated with $\Inert_\text{eff}$ [see Eq. \eqref{eq:InertEff}]:
\begin{equation}
|\eps_\text{eff}| \sim |\eps_P + \eps_\text{elastic} + \eps_\text{mag} + \eps_\parallel + \eps_\delta| \gtrsim |\eps_P|.
\end{equation}

\subsection{Evolution of magnetic inclination angle: analytic result for biaxial bodies}

Before presenting general results and applications in section \ref{sec:Crab}, we first summarize the key analytic results of \cite{Goldreich} for an effectively biaxial body $(e = 0,\infty)$.  We define three angles $\chi$, $\theta,$ and $\alpha$ by
\be
\cos\chi = \hat \vp \bcdot \ethreeeff , \;
\cos\theta = \hat \vomega \bcdot \ethreeeff, \;
\cos\alpha = \hat \vomega \bcdot \hat \vp.
\ee
We assume that the magnetic field axis $\hat \vp$ is frozen into the body, so that the angle $\chi$ is constant in time.  The other two angles in general will evolve.

On timescales much shorter than the pulsar spindown time, the variation of $\alpha$ is due to precession:
\be\label{eq:MagPrec}
\frac{\der \alpha}{\der t} \approx \omega_p \sin\chi \csc\alpha \sin\theta \sin(\omega_p t),
\ee
where $\omega_p = \cos\theta \, \eps_\text{eff} \, \omega$.  

Over timescales comparable to the pulsar the spindown time, the precession can be averaged out, giving \citep{Goldreich}:
\be
\frac{1}{\omega} \left\langle \frac{\der \omega}{\der t} \right\rangle \simeq - \frac{2 p^2 \omega^2}{3 c^3 I} \left[ \sin^2\chi + \sin^2\theta \left( 1 - \frac{3}{2} \sin^2\chi \right) \right],
\label{eq:Gold1}
\ee
\be
\frac{1}{\sin\theta} \left\langle \frac{\der \sin\theta}{\der t} \right\rangle \simeq - \frac{2 p^2 \omega^2}{3 c^3 I} \cos^2\theta\left( 1 - \frac{3}{2}\sin^2\chi \right).
\label{eq:Gold2}
\ee
From relation \eqref{eq:Gold2}, we see that $\theta$ always evolves toward $0^\circ$ or $90^\circ$, depending on if the angle $\chi$ is greater or less than the critical angle $\chi_\text{crit} = \sin^{-1}(\sqrt{2/3}) \simeq 55^\circ$.  This evolution takes place over the radiative timescale
\begin{multline}
\tau_\text{rad} = \frac{3 c^3 I}{2 p^2 \omega^2} = 7 \times 10^7  \\
\times \left( \frac{M}{1.4 \, M_\odot} \right) \left( \frac{P}{1 \, \text{s}} \right)^2 \left( \frac{R}{10^6 \, \text{cm}} \right)^{-4} \left( \frac{B_P}{10^{12} \, \text{G}} \right)^{-2}\text{years} .
\end{multline}

The magnetic inclination angle $\alpha$ is related to $\theta$ and $\phi = \omega_p t$ (the precession phase), and $\chi$, through the relation
\be\label{eq:cosalpha}
\cos\alpha = \hat \vomega \bcdot \hat \vp = \cos\chi \cos\theta + \sin\chi \sin\theta \cos\phi.
\ee
Using equation \eqref{eq:Gold2} and averaging over $\phi$, we obtain
\be\label{eq:MagAngle}
\left\langle \frac{\der \sin^2\alpha}{\der t} \right\rangle \simeq  - \frac{ p^2 \omega^2}{3 c^3 I} \sin^2( 2\theta) \left( 1 - \frac{3}{2}\sin^2\chi \right)^2.
\ee
This shows that $\alpha$ always decreases over timescales much longer than the precession period, but not necessarily to zero. From equation \eqref{eq:cosalpha}, we see if $\theta \to 0$, then $\alpha$ evolves to $\chi$, and if $\theta \to \pi/2$, then $\alpha$ evolves to $\pi/2 - \chi$.

\subsection{Counter-alignment of the Crab Pulsar}
\label{sec:Crab}

The Crab pulsar has $P = 0.0331 \, \text{s}$ and $\dot P = 4.22 \times 10^{-13}\text{s/s}$, implying the characteristic age of $P/2 \dot P = 1240 \, \text{years}$, and dipole field of order $B_P \sim 4 \times 10^{12} \, \text{G}$.  Through modeling the Crab pulsar pulse profile, many authors have estimated $\alpha$ to be in the range of $45^\circ - 70^\circ$ \citep{Harding(2008), Watters(2009), Du(2012)}.  The minimum effective ellipticity of the NS arises from the inertia of the dipole field, $\eps_P \sim 2.4 \times 10^{-12}$.  This would give a minimum precession frequency $\omega_p \gtrsim \eps_p \omega \sim 0.8^\circ/\text{year}$.  The observed $\der \alpha/\der t$ is $0.6^\circ/\text{century}$ \citep{Lyne(2013)}.  Since $\omega_p \gg \der\alpha/\der t$, to explain the observed $\der \alpha/\der t$ with precession [see Eq. \eqref{eq:MagPrec}], we require either $\chi \ll 1$ (the effective principal axis $\ethreeeff$ is almost aligned with the dipole axis) or $\theta \ll 1$ ($\ethreeeff$ is almost aligned with the rotation axis).  This would correspond to one of two special cases: the NS is dominated by stresses from the dipole magnetic field, or from rotation/ elasticity.

\begin{figure}
    \centering
    \includegraphics[scale=0.4]{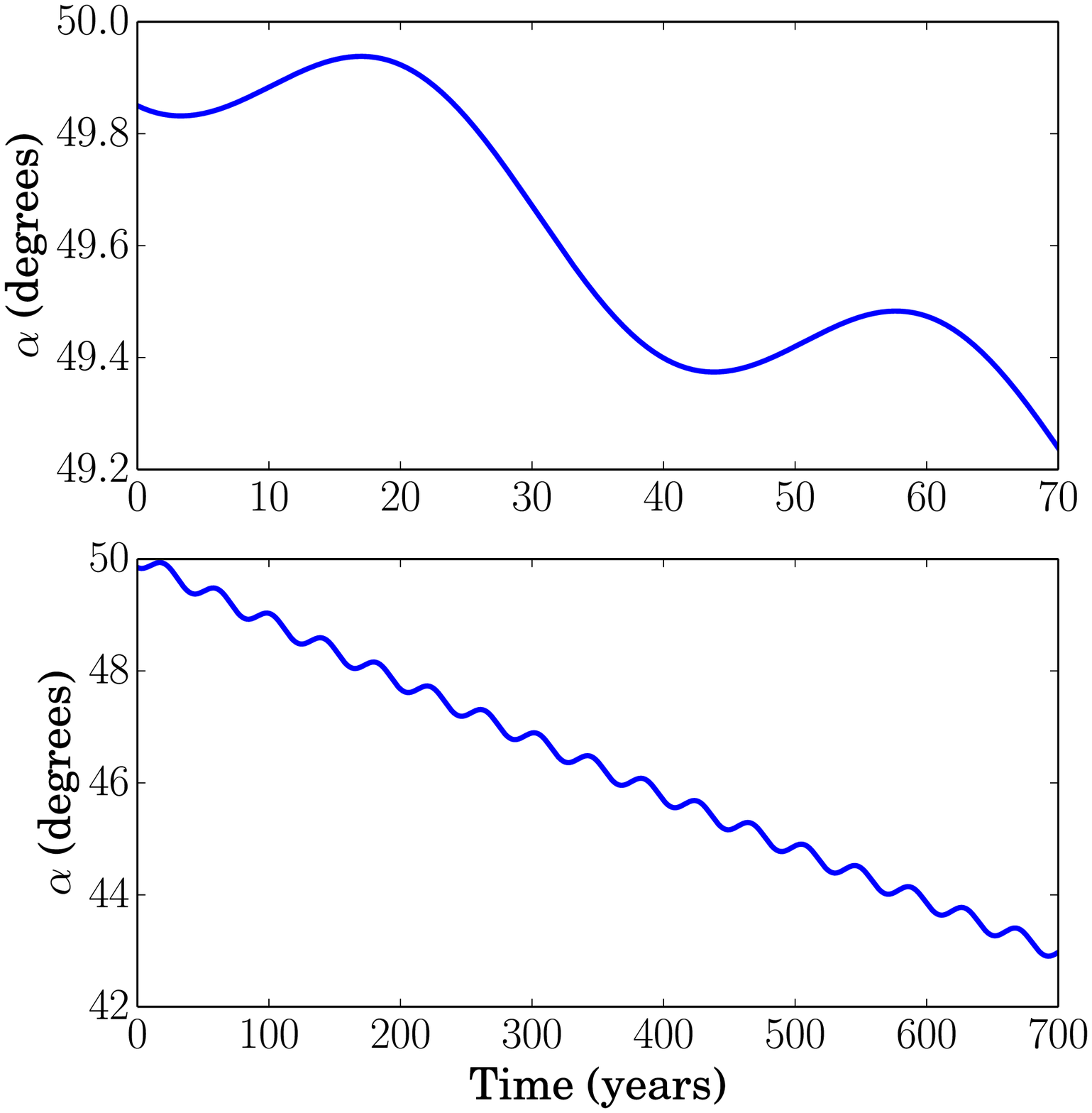}
    \caption{Evolution of magnetic inclination angle of a NS with an effectively biaxial $\Inert_\text{eff}$, with $\eps_\text{eff} = 4 \times 10^{-11}$.  The initial conditions are $\eoneeff \bcdot \hat \vomega = \sin(50^\circ)$, $\etwoeff \bcdot \hat \vomega = 0$, and $\ethreeeff \bcdot \hat \vomega = \cos(50^\circ)$.  The upper panel shows the variation over decade timescales.  We assume $\chi = 0.15^\circ$, which here is the angle between the dipole axis and the principal body axis $\ethreeeff$.}
    \label{fig:MagDom}
\end{figure}

\begin{figure}
    \centering
    \includegraphics[scale=0.4]{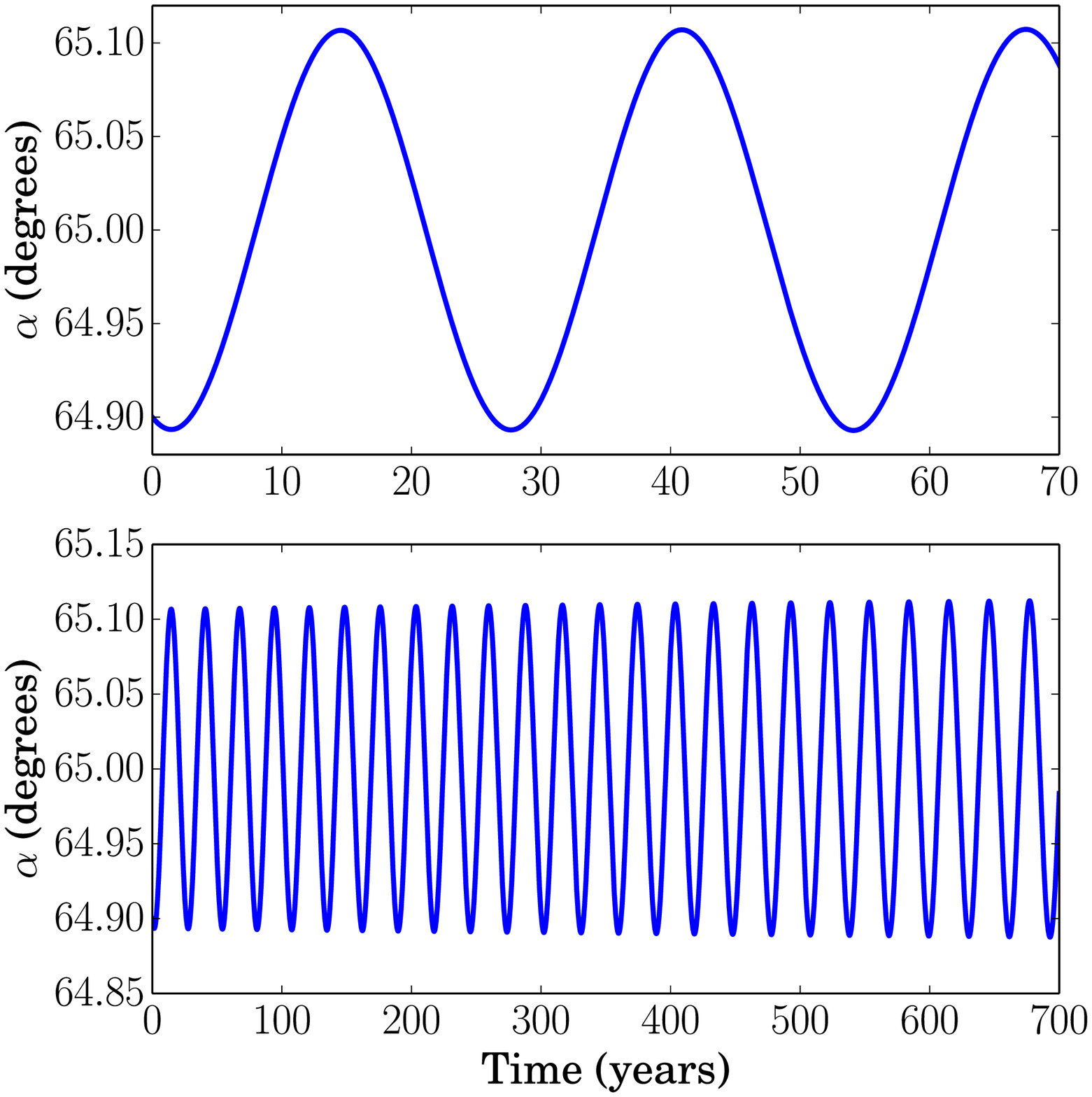}
    \caption{Same as Figure \ref{fig:MagDom}, except with $\chi = 65^\circ$ and an initial $\theta = 0.1^\circ$, which is the angle between the rotation axis and the principal body axis $\ethreeeff$.  In other words, the initial conditions are $\eoneeff \bcdot \hat \vomega(0) = \sin(0.1^\circ)\sin(65^\circ)$, $\etwoeff \bcdot \hat \vomega(0) = 0$, and $\ethreeeff \bcdot \hat \vomega(0) = \cos(0.1^\circ)\cos(65^\circ)$. }
    \label{fig:RotDom}
\end{figure}

Figures \ref{fig:MagDom} and \ref{fig:RotDom} depict two examples of the evolution of the magnetic inclination angle for a NS with an effectively biaxial $\Inert_\text{eff}$ and $\eps_\text{eff} = 4 \times 10^{-11}$.  This value could result from the elastic part of the rotational distortion [see \eqref{eq:epselastic}] or from the magnetic distortion associated with an internal field $B_*$ larger than the dipole field [see \eqref{eq:epsmag}].  Figure \ref{fig:MagDom} shows the case with $\chi = 0.15^\circ$, so that the principal axis $\ethreeeff$ is nearly aligned with $\hat \vp$, while Figure \ref{fig:RotDom} corresponds to the case with an initial $\theta = 0.1^\circ$, so that $\ethreeeff$ is nearly aligned with $\hat \vomega$.  In both cases, $\alpha$ increases during half of the precession phase, with $\dot \alpha$ consistent with the value $0.6^\circ/100 \, \text{years}$ observed by \cite{Lyne(2013)}.  

Figure \ref{fig:EffMagDom} depicts an example similar to Fig. \ref{fig:MagDom} except with $\eps_\text{eff} = 4 \times 10^{-11}$ and $\chi = 1.0^\circ$.  This value of $\eps_\text{eff}$ is close to the lower limit set by $\eps_P$ (associated with the inertia of the dipole field).  Note that with a mixture of comparable inertial poloidal and toroidal fields, the magnetic distortion $\eps_\text{mag}$ [Eq. \eqref{eq:epsmag}] can be reduced \citep{Mastrano(2013)}.  The rotational distortion $\eps_\text{elastic}$ [Eq. \eqref{eq:epselastic}] would not affect the precession if aligned with the spin axis.  Both sets of parameters (Fig. \ref{fig:MagDom} and \ref{fig:EffMagDom}) can account for the observed $\der \alpha/\der t$ of the Crab pulsar.  In the case of Figure \ref{fig:EffMagDom}, the $\dot \alpha > 0$ lasts for $\sim 200$ years because of the long precession period ($ 2\pi/\omega_p \propto \eps_\text{eff}^{-1}$),  whereas in the case of Figure \ref{fig:MagDom}, the $\dot \alpha > 0$ phase lasts $\sim 20$ years.

In both Figures \ref{fig:MagDom} and \ref{fig:EffMagDom}, we see a secular decrease of $\alpha$ over many precession cycles, with $\langle \dot \alpha \rangle \sim -1^\circ/100 \, \text{years}$.  Figure \ref{fig:RotDom} does not display a large secular change in $\alpha$.  This is because equations \eqref{eq:Gold1} and \eqref{eq:Gold2} predict $\theta$ stays close to zero, so $\alpha \sim \text{constant}$ according to equation \eqref{eq:MagAngle}.

\subsection{First and second order braking indexes}
\label{sec:break}

\begin{figure}
    \centering
    \includegraphics[scale=0.4]{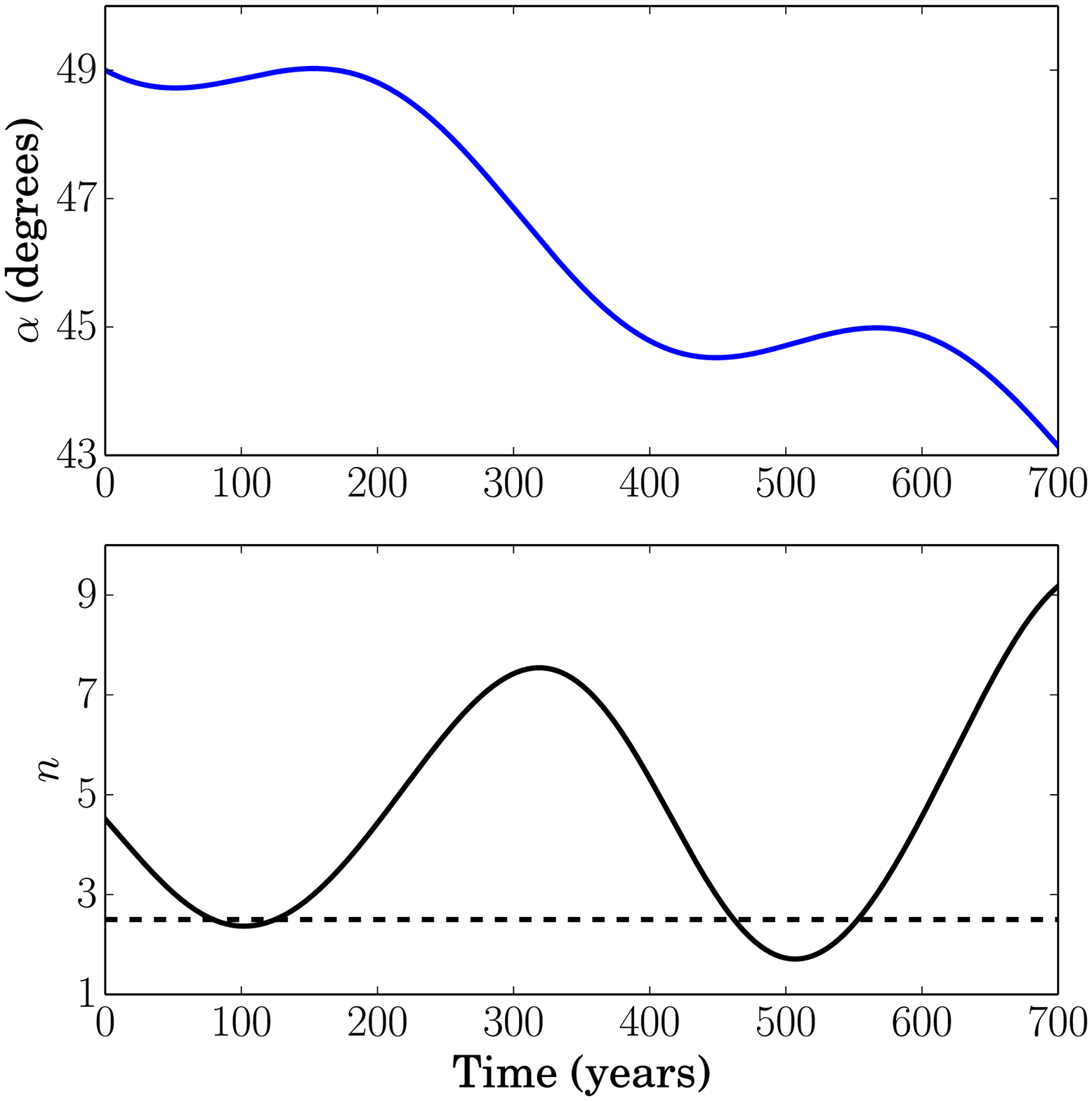}
    \caption{The evolution of magnetic inclination angle $\alpha$ (upper panel) and the braking index $n$ (lower panel).  The parameters are the same as Fig. \ref{fig:MagDom}, except with $\eps_\text{eff} = 4 \times 10^{-12}$ and $\chi = 1.0^\circ$.  The currently observed value $n = 2.5$ is indicated by the dashed line.}
    \label{fig:EffMagDom}
\end{figure}

Also plotted on Figure \ref{fig:EffMagDom} is the braking index $n~=~(\omega \ddot \omega)/\dot \omega^2$, computed using equation \eqref{eq:braking}.  We see that during the time when $\dot \alpha \sim 1^\circ/100 \, \text{years}$, the currently observed value of $n = 2.5$ is reproduced.  However, during the time when $\dot \alpha < 0$, a much larger value of $n$ is expected.  Thus continued observations of $n$ in the coming decades will test whether precession is responsible for the currently observed $\dot \alpha$ for the Crab pulsar.  

Modeling the dynamical evolution of the Crab pulsar through precession also gives predictions of the second order braking index, defined through $m \equiv (\dddot \omega \, \omega^2)/ \dot \omega^3$.  By re-writing in terms of $n$ and $\dot n$ \citep{MelatosCurrentStarved}, we have
\be
m = n(2n-1) + \frac{\omega}{\dot \omega} \dot n,
\ee
and using equations \eqref{eq:braking} and \eqref{eq:MagPrec},
\begin{multline}
m = n(2n-1) + (3-n)(n-1) \\
-2 (\dot \alpha)^2 \left( \frac{\omega}{\dot \omega} \right)^2(\cot^2 \alpha + \csc^2\alpha) + 2 \left( \frac{\omega}{\dot \omega} \right)^2 \dot \alpha \, \omega_P \, \cot(\omega_P t).
\end{multline}
With the observed $\dot \alpha = 0.6^\circ/\text{century}$ and $\omega/\dot\omega= -24.9 \, \text{century}$, and the estimate $\alpha \approx 45^\circ$ \citep{Harding(2008)}, we find
\be
m \approx 10.1 + 3.5 \left( \frac{\cos\theta}{1} \right) \left( \frac{\epsilon_\text{eff}}{10^{-11}} \right) \cot(\omega_P t).
\ee
If $\cot(\omega_P t) \approx 0$, we find precession gives an excellent agreement with the inferred 1993 value of $m \simeq 10.1$ for the Crab pulsar \citep{Lyne(1993)}.  An evolving precession phase ($\omega_P t$) may in part explain the increase of the second order breaking index to $m \simeq 45.6$ \citep{Lyne(2015)}.

We caution that our model does not include magnetospheric effects on the electromagnetic torque.  For example, if we adapt the spindown law \citep{Spitkovsky(2006)}
\be
\dot \omega \propto -\omega^3 (1 + \sin^2\alpha),
\ee
then equation \eqref{eq:braking} should be changed to
\be
n = 3 + 2 \frac{\omega}{\dot \omega} \frac{\sin \alpha \cos\alpha}{1 + \sin^2 \alpha} \dot \alpha.
\ee
Thus, to account for $n \simeq 2.50$ would require $\dot \alpha~\approx~1.7^\circ/100 \, \text{years}$, assuming $\alpha~\approx~ 45^\circ$.  The evolution of $n$ would be modified, along with the value of $m$.
  

\section{Discussion and Conclusion}
\label{sec:Conclusion}

In this paper, we have studied the rotational dynamics of magnetic
neutron stars (NSs), modeled as a non-spherical rigid body acted upon
by electromagnetic (EM) torques. First, we presented new
calculations of the near-field EM torques associated with the magnetic
inertia of the dipole and quadrupole fields of a NS in vacuum
[Eqs.~\eqref{eq:DipInertTorque} and \eqref{eq:QuadInertTorque}].  Our
analytical results show that if the NS has a quadrupole field a few
times stronger than the dipole field, the inertial quadrupole torque
can become more important than the corresponding dipole torque. Second,
we showed that, despite the complexity of the inertial torque
expressions, the effects of these
torques amount to a modification to the intrinsic moment of inertia tensor
of the star [see Eqs. \eqref{eq:Mp}-\eqref{eq:M} and \eqref{eq:InertEff}].
In general, the effective moment of inertia tensor is triaxial even
for an intrinsically biaxial NS. This allows us to understand
analytically the key effects of the inertial torques on the
precessional dynamics of magnetic NSs.  Finally, we applied our
theoretical results to the Crab pulsar in order to understand the
physical origin of the recently observed counter-alignment of the
pulsar's magnetic inclination angle $\alpha$ \citep{Lyne(2013),Lyne(2015)}.
We showed that it is possible to explain the increase of $\alpha$ on
decade timescales through precession.  However, since the typical
precession frequency $\omega_p$ is much greater than the observed rate
$\der \alpha/\der t$, this explanation requires some fine-tuning of
the principal axes $\{\hat \ve_{\text{eff},i} \}$ of the effective
moment of inertia tensor $\Inert_\text{eff}$: $\hat
\ve_{\text{eff},3}$ must be almost aligned with the magnetic dipole
axis or the spin axis.  This apparent fine-tuning may be expected if the star's intrinsic distortion arises primarily from the dipole field or from the rotation.  Over timescales comparable to
the pulsar's age, the magnetic inclination $\alpha$ always decreases
(see Figs. \ref{fig:MagDom}-\ref{fig:EffMagDom}). 

If the observed counter-alignment of magnetic inclination of the Crab
pulsar is indeed caused by precession, then the sign of
$\der\alpha/\der t$ will switch to negative after half a precession
period. Future observations would provide useful test and constrain 
the precession frequency (and thus the distortion of the pulsar).
Note that the upper bound on the time which the pulsar should take to reverse 
its counter-alignment behavior is $t \lesssim \eps_P^{-1} P/2 \sim 200 \;
\text{years}$ [see Eq. \eqref{eq:DipMag}].  Thus it is not unreasonable to suspect that the behavior of $\alpha$
may switch from counter-alignment to alignment within a human lifetime.

There are several complications and uncertainties neglected in our
model and calculations:
\begin{enumerate}
\item Our treatment of the crustal elasticity gave only an order of
  magnitude estimate of $\eps_\text{elastic}$, ignoring complications
  such as the NS equation of state and the thickness of the crust.
\item The internal magnetic field structure was mostly ignored,
  restricting our evaluation of $\eps_\text{mag}$ to an order of
  magnitude estimate.
\item We assumed that the only dissipative process was the radiative
  spindown torque.  Other dissipative processes such as polar
  wandering \citep{Macy(1974)} and crust-core couplings
  \citep{Shaham(1977), Alpar(1984)1, AlparOegelman(1987),
    CasiniMontemayor(1998), Sedrakian(1999), LinkCutler} may
  sigificantly affect the NS rotation/precession dynamics.
\item The exterior of the NS was assumed to be a vacuum, which is well
  known to not be true. The presence of a magnetosphere may change the
  inertial torques acting on the NS.
\end{enumerate}
Any one of these effects may affect the dynamical
evolution of $\alpha$, and further work is necessary to determine if
their inclusions will change the main results of this paper.

\section*{Acknowledgments}

We thank Andrew Melatos, Michael Kramer, Ira Wasserman, and James Cordes for useful
discussions, and the anonymous reviewer for quick and thoughtful suggestions improving the quality of our paper. This work has been supported in part by NSF grant
AST-1211061, and NASA grants NNX12AF85G, NNX14AG94G and NNX14AP31G.


\section{Appendix: EM torques on a rotating magnetic sphere in vacuum}

We begin with the magnetic field in the co-rotating frame of the star ``frozen in" to the body.  In the inertial frame, the magnetic field rigidly co-rotates with the star, maintaining the same shape and magnitudes as in the rotating frame.  If the star is perfectly conducting, the electric field inside the star in the inertial frame is given \textit{exactly} by $\E = -(\vv/c) \btimes \B$ \citep{LLEM}.  If we work in the spherical coordinates $(r,\theta,\phi)$, the coordinate transformation from the co-rotating frame to the inertial frame is $\phi \to \phi-\omega t$, where $\omega$ is the rotation rate.  If the surface of the star is given by $r=R$, the electromagnetic fields must satisfy the boundary conditions $(\br \bcdot \B)_{r=R+} = (\br \bcdot \B)_{r=R-}$ and $(\br \btimes \E)_{r=R-} = (\br \btimes \E)_{R=R+}$.  Thus, the external EM fields of the rotating star are uniquely determined by the normal components of the magnetic field at the stellar surface.

The complete solution of the EM fields in vacuum in terms of multipole moments is derived in \cite{Jackson}.  These fields, seperated into time-independent terms $\B_0$ and $\E_0$, and time-dependent terms $\E'$ and $\B'$, may be re-expressed in terms of vector spherical harmonics:
\begin{align}
\B_0 = \sum_\ell &\left[ (\ell+1) a_M(\ell,0) \left( \frac{\xo}{x} \right)^{\ell+2} \vY_{\ell 0} \right. \nonumber\\
&- \left. a_M(\ell,0) \left( \frac{\xo}{x} \right)^{\ell+2} \vPsi_{\ell 0} \right], \\
\E_0 = \sum_\ell &\left[ (\ell+1) a_E(\ell,0) \left( \frac{\xo}{x} \right)^{\ell+2} \vY_{\ell 0} \right. \nonumber \\
&- \left. a_E(\ell,0) \left( \frac{\xo}{x} \right)^{\ell+2} \vPsi_{\ell 0}\right], \\
\B' = \sum_{\ell,m\ne0} &\left\{ -\im \frac{a_E(\ell,m)}{\sqrt{\ell(\ell+1)}} h_\ell(m x) \vPhi_{\ell m} \right. \nonumber \\
&+ \left. \sqrt{\ell(\ell+1)} a_M(\ell,m) \frac{h_\ell(m x)}{m x} \vY_{\ell m} \right. \nonumber\\
&+ \left. \frac{a_M(\ell,m)}{\sqrt{\ell(\ell+1)}} \left[ \frac{h_\ell(m x) + mxh'_\ell(mx)}{mx} \right] \vPsi_{\ell m} \right\}, \\
\E' = \sum_{\ell,m\ne0} &\left\{ -\sqrt{\ell(\ell+1)} a_E(\ell,m) \frac{h_\ell(mx)}{mx} \vY_{\ell m} \right. \nonumber \\
 &- \left.\frac{a_E(\ell,m)}{\sqrt{\ell(\ell+1)}} \left[ \frac{h_\ell(mx)+mxh'_\ell(mx)}{mx} \right] \vPsi_{\ell m} \right. \nonumber \\
 &- \left. \im \frac{a_M(\ell,m)}{\sqrt{\ell(\ell+1)}} h_\ell(mx) \vPhi_{\ell m} \right\},
\end{align}
where
\be
\begin{array}{cl}
 \vY_{\ell m} \equiv & Y_{\ell m}(\theta,\phi-\omega t) \hat \br, \\
 \hspace{2mm} \vPsi_{\ell m} \equiv & r \vdel Y_{\ell m}(\theta,\phi - \omega t), \\
\hspace{2mm} \vPhi_{\ell m} \equiv & \br \btimes \vdel Y_{\ell m}(\theta,\phi - \omega t),
\end{array}
\ee
with $x \equiv r\omega/c$ and $Y_{\ell m}$ denote spherical harmonics.  The actual fields are given by the real parts of $\E$ and $\B$.  Defining $q_{\ell m} \equiv \int \der \Omega (\B \cdot \hat \br)_{r=R}$, the boundary conditions give the multi-pole moments:
\begin{align}
a_M(\ell,m) &= \left\{
\begin{array}{ll}
\frac{q_{\ell 0}}{\ell+1} & m=0 \\
\frac{q_{\ell m}}{\sqrt{\ell(\ell+1)}} \frac{m\xo}{h_\ell(m\xo)} & m\ne 0
\end{array} \right. \\
a_E(\ell,m) &= \left\{
\begin{array}{ll}
\frac{\xo}{\ell(\ell+1)} \left[  (1-\delta_{\ell 0})\ell J_{(\ell+1)0}q_{(\ell+1)0} \right. & {} \\
\left. \hspace{10mm} -  (\ell+1) J_{\ell 0} q_{(\ell-1)0} \right] & m=0 \\
\frac{\xo}{\sqrt{\ell(\ell+1)}} \frac{m\xo}{h_\ell(m\xo) + m\xo h_\ell'(m\xo)} & {} \\
\times \left[  \ell J_{(\ell+1)m} q_{(\ell+1)m} \right. & {} \\
\left. -  (\ell+1)J_{\ell m} q_{(\ell-1)m}  \right] & m\ne 0
\end{array} \right.
\end{align}
where $\xo \equiv R\omega/c$, $\delta_{\ell m}$ is the Kronecker delta, and
\be
J_{\ell m} \equiv 
\left\{ \begin{array}{cc}
\sqrt{ \frac{\ell^2 - m^2}{4l^2 - 1} } & \text{if} \; |m|<\ell \\
0 & \text{if} \; |m| \ge \ell
\end{array} \right.
\ee
The Kronecker delta term forces charge neutrality.

It may be shown that for $\ell = 1$ the above expressions reproduce the solution by \cite{Deutsch(1955)} [see \cite{MelatosCurrentStarved} for corrections].  For this work, we are interested in both the $\ell = 1$ and $\ell = 2$ terms.  Let the unit vector in the direction of the dipole moment be $\hat \vp$, and the three unit eigenvectors for the quadrupole moment be $\qone$, $\qtwo$, and $\qthree$ (see Fig \ref{fig:Quad_field}).  If the magnetic field consists only of $\ell = 1$ term, we may define the magnitude of $B_P$ as $B_P = B(R, \hat \vp)$, where the notation $B(r,\hat {\bf n})$ denotes the magnetic field evaluated at $r$ with angular coordinates in the direction of the unit vector $\hat {\bf n}$.  The quadrupole magnetic field has two independent components, due to the fact that the quadrupole tensor $\mathbf{Q}$ is trace free (see Fig. \ref{fig:Quad_field}).  We define these to be $B_\parallel \equiv B(R,\qone)$ and $B_\delta \equiv [B(R,\qthree)-B(R,\qtwo)]/2$.  

The time-averaged EM Maxwell stress tensor is given by
\be
\T = \frac{1}{4\pi}\text{Re}\left[\E \otimes \E^* + \B \otimes \B^* - \frac{1}{2}\one(|E|^2 + |B|^2) \right],
\ee
where $\one$ is the unit dyadic, and the stars denote the complex conjugate.  One may decompose $\T$ into spherical coordinates to evaluate the torque $\vGamma$ in Cartesian coordinates:
\be\label{eq:TorqueComps}
\begin{array}{cl}
\Gamma_x = & -R \int \der\Omega \left[ T_{r\theta} \sin\phi + T_{r\phi} \cos\theta \cos\phi \right]_{r=R}, \\
\Gamma_y = & R \int \der\Omega \left[ T_{r\theta} \cos\phi - T_{r\phi} \cos\theta \sin\phi \right]_{r=R}, \\
\Gamma_z = & R \int \der \Omega \left[ T_{r\phi} \sin\theta \right]_{r=R},
\end{array}
\ee
where $x + \im y = r \sin\theta e^{\im \phi}$, $z = r\cos\theta$.

\textit{After} integrating equation \eqref{eq:TorqueComps}, we expand the expressions in $\xo$ and only keep terms proportional to $\xo^2$.  We then arrive at equations \eqref{eq:DipInertTorque} and \eqref{eq:QuadInertTorque}.

\end{document}